# Ball Lightning as a Self-Organized Complexity

**Erzilia Lozneanu, Sebastian Popescu** and **Mircea Sanduloviciu**
Department of Plasma Physics
"Al. I. Cuza" University
6600 Iasi, Romania
msandu@uaic.ro

The ball lightning phenomenon is explained in the frame of a self-organization scenario suggested by experiments performed on the spontaneously generated complex spherical space charge configuration in plasma. Originated in a hot plasma, suddenly created in a point where a lightning flash strikes the Earth surface, the ball lightning appearance proves the ability of nature to generate, by self-organization, complex structures able to ensure their own existence by exchange of matter and energy with the surroundings. Their subsequent evolution depends on the environment where they are born. Under contemporary Earth conditions the lifetime of such complexities is relatively short. A similar self-organization mechanism, produced by simple sparks in the earl Earth's atmosphere (chemically reactive plasma) is suggested to explain the genesis of complexities able to evolve into prebiotic structures.

## 1. Introduction

The enigmatic ball lightning phenomenon has stimulated the people interest for a very long time. Thus the ball lightning appearance was noted with more than a hundred years ago in prestigious periodicals as for example "Nature". The vast literature on ball lightning comprises books entirely dedicated to this subject [1-3], review papers focussed on this topic [4-7], as well as data banks. The data banks contain collections of characteristics for lightning balls that where occasionally observed under different circumstances. For example, in the United States, lightning balls have been observed by about 5% of the adult population at some time of their lives [8]. Among the prominent individuals who have observed lightning balls are Niels Bohr and Victor Weisskopf [9]. Periodically, new aspects concerning the ball lightning are discussed at international symposiums, the last one taking place in 1999 at Antwerp, in Belgium.

The intriguing appearance and characteristics of the ball lightning have given rise to a wide variety of models proposed for explaining the phenomenon. At present the scientists that have focused their investigations on the ball lightning could be divided in two parts. One part considers that the plasma, created by an ordinary lightning stroke in the well-localized region where it strikes the Earth surface, contains a sufficient amount of matter and energy to explain the ball lightning appearance and lifetime. This energy could eventually be stored as chemical one in a special kind of accumulator [10], or is related to a more complex chemical phenomenon involving, for example, water vapors [6]. Also other forms of energy, as for example nuclear energy [11] and processes related to the behavior of antimatter [12] were occasionally invoked to be at the origin of ball lightning. The energy content and, consequently, the lifetime of ball lightning are evidently related (in all of these models) to the amount of energy transferred from the ordinary linear lightning to the ball lightning. After this energy transfer, the ball lightning appears, for a certain time sequence, as an electrical gas discharge [10] or as a result of possible other reactions (as for example chemical ones) accompanied by light emission [6]. A new ball lightning model, published in Nature [13], considers at the origin of ball lightning the chemical energy stored in nanoparticles of silica-carbon mixtures which are ejected into the air as a filamentary network when a normal lightning strikes the soil.

Another part of scientists relate the ball lightning appearance to the presence of electromagnetic phenomena that usually accompany electrical discharges, as for example ordinary linear lightning produced in the Earth atmosphere during stormy weather, but occasionally also under fair weather conditions. The prominent scientist that has proposed such a model was P. L. Kapitsa [14]. He has proved the possibility to create in laboratory, at atmospheric pressure, by interference, high frequency discharges, well localized, showing some similarities with ball lightning. Afterwards the ball lightning model proposed by Kapitsa was justified by theoretical considerations based on the presumption that electromagnetic radiation is trapped in a plasma shell formed when a large amount of radiant energy expels the plasma by ponderomotive force [15]. A similar theory relating the ball lightning appearance to radiant energy produced during stormy weather was proposed by Endean [16].

## 2.    On the implications of self-organization in the ball lightning phenomenology

The possibility to explain the appearance of the ball lightning by self-organization was firstly suggested by one of us at an international symposium [17] in 1988, and then extensively described in other papers [18-21]. Three years later Kadomtsev [22] suggested the considering of ball lightning as a result of a self-organization mechanism presumably possible in cold dusty plasma with electrochemical active particles.

The arguments for considering the self-organization as the genuine origin of the ball lightning were inspired by experiments performed by us in different plasma devices. As is well known, plasma is a medium where self-confined space charge structures appear as flaming globes. These structures have been observed for long time ago. Thus, in 1920s the American Nobel laureate Irving Langmuir, considered the "father" of Plasma Physics, described an experiment in which he obtained luminous globes that seem to have characteristics similar to those ascribed to ball lightning [23]. Over the years, the spontaneous formation of spatial and spatiotemporal patterns, frequently observed in plasma, have been intensively studied both by experimental and theoretical physicists. These studies have emphasized striking similarities of these patterns with those formed in solid state physics, but also in chemical and biological media [24,25]. Such similarities justify the opinion of considering the ball lightning problem as one potentially solvable in the frame of nonequilibrium physics. Arguments for this attempt are presented in a recently published paper [21] that contains a test experiment able to reveal the nature of physical processes whose successive appearance in a plasma, subjected to an external constraint, determines the creation of a well defined flaming globe. The succession of physical processes experimentally identified in this test experiment reveals the presence of a self-organization scenario potentially possible also in the Earth atmosphere in the presence of electrical activity. The described experiment tries to join the hypotheses dominant today concerning the ball lightning genesis. First, there is the hypothesis that the energy injected in the ball lightning by an ordinary linear lightning is itself sufficient for explaining the ball lightning appearance and lifetime. Second, there is the hypothesis that the occasionally observed long lifetime of ball lightning is ensured by trapping RF energy from external sources, i.e. lightning discharges produced in the Earth atmosphere during stormy weather. This test experiment has justified our attempt to consider it as a laboratory simulation of ball lightning. With its help we have established a direct relationship between the nonlinear behavior of plasma subjected to an external constraint and the key processes generally accepted to be relevant for the presence of self-organization. As is well known such key processes are: i) symmetry breaking, ii) structural instability, iii) bifurcation, and iv) long-range order [26]. The presence of the above mentioned key processes was proved experimentally by gradually injecting matter and energy in a plasma conductor placed between two electrodes. Working as a plasma diode, the used experimental device makes possible to reveal abrupt changes of the electrical conductivity of plasma by plotting the current versus voltage static characteristic. Investigating the causes that explain the observed abrupt changes of the plasma conductivity (emphasized by the presence of critical points in the static characteristic) we have obtained information concerning the genuine origin of the above listed key processes. In this way we identified the stages of the intermittent scenario of self-organization, whose final product is a flaming globe with behavior similar to that of a ball lightning. Although not possible under natural conditions in the Earth atmosphere, the identified scenario is essential for understanding the self-assemblage process of the ball lightning considered as a self-organized complex space charge configuration (CSCC). Based on such a self-organization scenario, it becomes possible to explain the appearance of the ball lightning by considering as initial condition the creation of a hot plasma in the point where an ordinary lightning strikes the Earth surface. The subsequent evolution of the plasma produced by local vaporization of soil can be described in the frame of the so-called cascading self-organization process [27]. Such a cascading scenario of self-organization is appropriate to the conditions under which ball lightning appears, because the sudden injection of matter and energy in the form of ordinary lightning is a process evidently present in nature. In comparison with the experiment performed in a plasma diode, largely described in [21], where the plasma was created before the matter and energy were gradually injected, under stormy weather conditions the lightning stroke creates itself a hot plasma in the point where it strikes Earth's surface.

The evolution of the hot plasma towards a free-floating flaming globe, namely the ball lightning, becomes possible only if certain conditions are satisfied. The first condition requires the spatial separation of electrons and positive ions from the hot plasma. In the phase when the hot plasma is in electrical contact with the positively charged Earth, such a phenomenon becomes possible taking into account the high mobility of electrons with respect to that of positive ions. As a consequence, the Earth quickly collects electrons, so that a nucleus enriched in positive ions appears. Because of the high temperature of the plasma created by the lightning stroke, the separation of opposite electrical space charges is enhanced by thermal diffusion, so that the positive potential of the nucleus becomes higher than that of the Earth [21]. The assemblage of the ball lightning can be explained considering the following succession of sequential steps related to the key processes of self-organization (specified in the following by italics):

1. Creation of a well localized hot plasma at the impact point of an electrical discharge, most frequently an ordinary linear lightning - *initial external constraint.*

2. Separation of electrical opposite space charges because of the differences in the diffusivity and mobility of electrons and positive ions- *nonequilibrium state.* A nucleus enriched in positive ions, surrounded by a partially ionized gas that fades off into the neutral atmosphere, appears as a final product.

3. Acceleration towards the positive nucleus of the surrounding thermalized electrons up to energies for which the neutral excitation and ionization cross section functions suddenly increase in two adjacent regions - *symmetry breaking and spatial separation of the system's functions relevant for self-organization.*

4. Accumulation of those electrons that have lost their momentum after neutrals' excitation in the region where the cross-section function of this process suddenly increases. Development of electrostatic forces acting as long-range correlations between the net negative space charge formed after electrons' accumulation and the positive ions present in the nucleus - *initiation of long-range order.*

5. Accumulation of opposite space charges in the form of a nearly spherical electrical double layer that surrounds the positive nucleus. In this phase the "fireball" is in electrical contact with the Earth. Sudden transition into a state characterized by local minimum of the free energy that spontaneously appears by the transition of the double layer into a spherical one- *structural instability.* The enclosed spherical space charge configuration simultaneously detaches from the Earth surface. So, a free-floating flaming globe, namely the ball lightning, appears as a relatively stable self-organized complexity.

6. Ensurance of the ball lightning existence for a certain time span by a proper dynamics and for a certain time span, by a periodical exchange of matter and energy with the surrounding plasma mantle. This behavior explains the sound and electromagnetic phenomena reported as observational characteristics of ball lightning.

The above-mentioned detachment of the ball lightning from the Earth surface depends sensitively on the conductivity of the soil in the point where the linear lightning produces the hot plasma. When the soil conductivity is sufficiently low, a part of electrons collected by the Earth, during the positive charging process of the nucleus, remains in the vicinity of the contact point of the nearly spherical space charge configuration with the Earth surface. When the space-charge transits into a configuration characterized by a minimum of the free energy, the nearly spherical double layer becomes closed. In this moment repulsive electrostatic forces acting between the external side of the spherical double layer and the negative space charge present for a short time span at the Earth surface determine the detachment process. The simultaneous fulfillment of the above both described processes explains the very rare appearance of the ball lightning in the Earth atmosphere. In the further evolution, the net negative charge well localized on the Earth surface disperses so that the balance between the attractive electrostatic forces and the buoyancy governs, for a relatively long time, the ball lightning position and displacement with respect to the Earth surface.

Based on the described scenario, the ball lightning could be considered as a giant "cell" created by self-organization in the contemporary Earth atmosphere [21].

## 3. Explanation of the ball lightning observational characteristics

In practice, all observational characteristics of the ball lightning listed in [6] can be explained considering it as a self-organized CSCC [21]. Thus, its association with stormy weather is the premise for creation of the hot plasma by a lightning stroke. Its emission of light is related to the de-excitations of the neutrals that were excited by electrons accelerated towards the positive nucleus. The spherical shape corresponds to a space charge configuration characterized by a minimum of the free energy. Therefore the ball lightning self-assemblage, started from an initially hot plasma, does not require expenditure of additional external energy. Only internal processes related to the described cascading self-organization scenario govern the subsequent evolution of the hot plasma into a lightning ball. The different colors of lightning balls and the associated smell could be explained considering the nature of materials evaporated in the point where the lightning stoke strikes the Earth surface. The ball lightning's size depends on the amount of energy transferred by the ordinary lightning that has created the hot plasma but also, after its creation, on its ability to act as a cavity able to absorb at resonance radiant energy. The considerable stability as an entity could be related to the presence of a spherical double layer that ensures the self-confinement mechanism emphasized by the spatial coherence of the ball lightning. The apparent coldness of the ball lightning has a plausible explanation knowing that the energy required for the self-assemblage of an electrical double layer has the order of magnitude the neutrals' ionization energy. The lack of buoyancy of the ball lightning is related to its gaseous nature self-confined by an electrical double layer. Its floating state, realized after its detachment from the Earth surface, is explainable considering the presence of attractive electrostatic forces acting between the net negative space charge localized at the external side of the double layer and the positive Earth. The buoyancy force equilibrates those attractive forces to which the gravitational one must also be added. A plausible explanation of the occasionally observed erratic motion of the ball lightning, also against the wind, necessarily requires the consideration of the transition from a stationary double layer into a moving one. This transition takes place when a local self-enhancement of the production of positive ions determines the overcompensation of the adjacent net negative space charge accumulated after neutrals' excitation. The two modes of demise, explosively or simple disappearing depend on the energy transferred to the initial plasma that generates the ball lightning, but also on the radiant energy absorption in the moment of its de-aggregation.

Another important observational characteristic is the occasionally observed penetration through dielectric walls without producing any damage in the material [4, 28]. Such a phenomenon becomes possible considering the ball lightning as a cavity that absorbs at resonance high frequency radiant energy. Under such conditions the apparent penetration, for example, through a window of an aircraft [4] is, very probable, a re-generation of the ball lightning at the other side of the dielectric wall.

## 4.  Cell-like characteristics of the ball lightning

From the point of view of nonequilibrium physics one of the most interesting observational characteristics of the ball lightning is, beside its genesis, its transition into a state in which it emits sound and electromagnetic waves. This proves that the ball lightning naturally evolves, immediately after its "birth", into a state in which a rhythmic exchange of energy and matter with the surrounding plasma mantle ensures its own existence for a relatively short time span. The magnitude of this time span depends on the amount of slow electrons stored in the mantle. An additional heating process, ensured by absorption of radiant energy from eventually present RF sources, can extend the lifetime of the ball lightning. The matter and energy exchange process, sustained and controlled by the double layer that protects the CSCC like a cell membrane, makes the ball lightning phenomenology even more interesting for scientists whose goal is to elucidate the mechanism by which the Nature has created the first pre-biotic complex structure. Such structures might have appeared under pre-biotic Earth's conditions when usual electrical sparks had created, in the above-described way, a CSCC. In contrast to the present Earth conditions, the CSCC had been self-assembled in an environment (reactive plasma) in which a subsequent evolution into o more complex, perhaps pre-biotic, structure became potentially possible. In this context it is interesting to remind the content of already published papers that have reported the appearance of "micro lightning balls" in plasma devices [29], but also the self-assemblage, in reactive HF plasma, of membranous vesicles [30-33]. These vesicles are considered as an essential step in the evolution of the earliest cells. Some remarkable behavior of the CSCC formed in laboratory, as for example its self-replication by division, but also its ability to transmit information by radiant energy (related to the double layer dynamics), absorbed at resonance by similar other CSCC, put forward some qualities usually attributed only to living systems [30].

Evidently, the attempt to explain the origin of life, starting from a gaseous cell self-assembled in the chemical reactive early Earth atmosphere by a scenario of self-organization similar to that creating lightning balls in the contemporary atmosphere must be approached with great care. Additional laboratory experiments concerning the evolution of a gaseous cell-like configuration, created by a usual electrical spark produced in suitable selected reactive plasmas, into a vesicle, could eventually elucidate this striking problem of broad interest for scientists working in the most interdisciplinary science, namely the Science of Complexity.

## 5. Conclusions

Based on new knowledge concerning the scenario of self-organization that explains the appearance of CSCC in plasmas we have described in this paper a phenomenological model of the ball lightning. This model suggests that a local injection of matter and energy by a lightning stroke can start, in certain conditions, the self-assemblage of a well defined free floating flaming globe, named ball lightning. The same self-organization scenario initiated by simple sparks under prebiotic Earth conditions is considered as a possible answer to the problem concerning the manner by which the Nature has created initially gaseous CSCCs having the ability to evolve into precursors of living complex systems.